\documentclass[10pt,preprint]{aastex}
 \usepackage{natbib}
\def\FF{\mathcal{F}}
\def\ang{\AA}

\def\gapprox{\lower.4ex\hbox{$\;\buildrel >\over{\scriptstyle\sim}\;$}}
\def\lapprox{\lower.4ex\hbox{$\;\buildrel <\over{\scriptstyle\sim}\;$}}

\shortauthors{Aschwanden 2022}
\shorttitle{Universality of Power Law Slopes: HMI and IRIS}

\begin{document}
\renewcommand{\topfraction}{0.95}
\renewcommand{\bottomfraction}{0.95}
\renewcommand{\textfraction}{0.05}
\renewcommand{\floatpagefraction}{0.95}
\renewcommand{\dbltopfraction}{0.95}
\renewcommand{\dblfloatpagefraction}{0.95}


\title{ The Universality of Power Law Slopes 
	in the Solar Photosphere and Transition Region
	Observed with HMI and IRIS}

\author{Markus J. Aschwanden}
\affil{Lockheed Martin, Solar and Astrophysics Laboratory (LMSAL),
       Advanced Technology Center (ATC),
       A021S, Bldg.252, 3251 Hanover St.,
       Palo Alto, CA 94304, USA;
       e-mail: aschwanden@lmsal.com}

\and 
\author{Nived Vilangot Nhalil}
\affil{Armagh Observatory and Planetarium, College Hill, Armagh BT61 9DG, UK}

\begin{abstract}
We compare the size distributions of {\sl self-organized criticality (SOC)}
systems in the solar photosphere and the transition region, using magnetogram
data from {\sl Helioseismic and Magnetic Imager (HMI)} and {\sl Interface 
Region Imaging Spectrograph (IRIS)} data. For each dataset we fit a
combination of a Gaussian and a power law size distribution function, 
which yields information on four different physical processes: 
(i) Gaussian random noise in IRIS data;
(ii) spicular events in the plages of the transition region 
(described by power law size distribution in IRIS data);
(iii) salt-and-pepper small-scale magnetic structures
(described by the random noise in HMI magnetograms);
and (iv) magnetic reconnection processes in flares and nanoflares
(described by power law size distributions in HMI data).
We find a high correlation (CCC=0.90) between IRIS and HMI data.
Datasets with magnetic flux balance are generally found to 
match the SOC-predicted power law slope $\alpha_F=1.80$ 
(for mean fluxes $F$), but exceptions occur due to arbitrary
choices of the HMI field-of-view. The matching cases
confirm the universality of SOC-inferred flux size distributions, 
and agree with the results of Parnell et al.~(2009), 
$\alpha_F=1.85\pm0.14$.
\end{abstract}


\section{	INTRODUCTION 		}  

 	Self-Organized Criticality (SOC)
	is a critical state of a nonlinear energy 
        dissipation system that is slowly and continuously 
        driven towards a critical value of a system-wide 
        instability threshold, producing scale-free, 
        fractal-diffusive, and intermittent avalanches 
        with power law-like size distributions  
        (Aschwanden 2011). The original
        paradigm and characteristic behavior of SOC 
        systems was studied from sandpile avalanches,
        based on the next-neighbor interactions in 
        microscopic lattice grids (Bak et al.~1987, 1988),
	also called cellular automaton algorithms.
        However, alternative macroscopic models can mimic the
        same system behavior also, based on macroscopic
        power law scaling laws of correlated physical 
        parameters. For instance, the hard X-ray flux
        radiated in a solar flare was found to 
        scale with the (fractal) spatial volume of the flare.
        The exponentially growing instability that produces
        the flare predicts a well-defined power law 
	size distribution function,  
        which applies also to a host of other nonlinear
        systems, such as earthquakes or stock
        market fluctuations, in contrast to linear systems,
        such as Gaussian noise. Thus, modeling of SOC systems
        helps us to discriminate between linear
        and nonlinear systems. Knowledge of the correct
        size distributions yields us statistical predictions 
        of the largest catastrophic events in SOC systems. 
	Besides flaring and heating of the solar corona, 
	we hope to obtain also new insights into nanoflaring 
	in the atmosphere of the Quiet Sun, which we pursue here.

The atmospheric structure of the Sun consists of the photospheric
layer on the solar surface, the chromosphere, the transition region,
the corona, and solar wind regions, which all host different physical
processes, characterized by the electron density, the electron 
temperature, and the magnetic field strength. In this study we 
sample very diverse temperature structures, from $T_e \approx 5800$ K
observed in photospheric magnetograms with the {\sl Helioseismic 
and Magnetic Imager (HMI)}, to $T_e \approx 10^4-10^5$ K, observed
in Slitjaw images (SJI) of the 1400 \ang\ channel of IRIS,
which are dominated by the Si IV 1394 \ang\ and 1403 \ang\
resonance line, and form in the transition region
(Rathore and Carlsson 2015; Rathore et al.~2015).
Due to this huge temperature range, different physical processes
are dominant in the various temperature regimes
(Gallagher et al.~1998; Warren et al.~2016), and thus we do
not know {\sl a priori} whether the concept of {\sl self-organized criticality 
(SOC)} systems (Aschwanden 2011; 2014; Aschwanden et al.~2016;
McAteer et al.~2016; Warren et al.~2016) is applicable. More specifically, 
we want to understand the functional shapes of observed occurrence frequency 
(size) distributions, and whether they exhibit power law function (slopes) 
with universal validity in different temperature and wavelength regimes.

There is an ongoing debate on the functional form of size distributions
in avalanching SOC processes, such as: 
a power law function, 
a log-normal distribution (Verbeeck et al.~2019), 
a Pareto distribution (Hosking and Wallis 1987), 
a Lomax distribution (Lomax 1954; Giles et al.~2011),
or a Weibull distribution (Weibull 1951), for instance. 
Since all these 
functional forms are close to a power law function on the right-hand 
side of the size distribution, which is also called the ``fat-tail'',  
various linear combinations of these functional forms have been
found to fit the observed size distributions with comparable accuracy  
(Munoz-Jaramillo et al.~2015). In this study we use a combination of
(Gaussian) incoherent random and (power law-like) coherent random structures.
Gaussian statistics reflect the
operation of a memoryless stationary (incoherent) random process; while
avalanching (coherent) processes such as occurring in SOC systems are
characterized by extended spatial and temporal correlations
(i.e., the unfolding of an avalanche is influenced by the
imprint of earlier avalanches on the system; see Jensen 1998, chapter 2).

Here, the incoherent  component describes the Gaussian noise 
(visible in IRIS data), as well as the salt-and-pepper structure 
(visible in HMI magnetograms). On the other side, the coherent noise of the  
power law component may be produced by the spicular dynamics 
(visible in IRIS data), or by magnetic reconnection dynamics 
of small-scale features and nanoflares (visible in HMI magnetograms).
Gaussian noise distributions have been tested with Yohkoh soft X-ray data
(Katsukawa and Tsuneta 2001).
Log-normal distributions, which are closest to our Gaussian-plus-power-law
method used here, have been previously studied for 
Quiet-Sun FUV emission (Fontenla et al.~2007),
solar flares (Verbeeck et al. 2019),
the solar wind (Burlaga and Lazarus 2000),
accretion disks (Kunjaya et al.~2011), 
and are discussed also in Ceva and Luzuriaga (1998), 
Mitzenmacher (2004), 
and Scargle (2020).

A new aspect of this study is the invention of a single-image
algorithm to derive ``pixelized'' size distributions
$N(F) \propto F^{-\alpha_F}$. A major test consists of
comparing the observed power law slopes $\alpha_F$ with the 
theoretical SOC-predicted values. 
Another crucial test is the power law slope $\alpha_E$ of nanoflare 
energies, which is decisive for testing the coronal 
heating energetics (Hudson 1991; Krucker and Benz 1998;
Vilangot Nhalil et al.~2020; Aschwanden 2022b). 
Numerous studies have inferred SOC parameter
correlations of impulsive events in the outer solar 
atmosphere, in an attempt to understand the predominant 
energy supply mechanism in the corona (Vilangot Nhalil 
et al.~2020), which motivates us to pursue a follow-on 
study, using data from sunspots and plages to further 
investigate bright impulsive events in the transition region.
Ultimately, we strive for a unification of small-scale 
phenomena in the solar corona and transition region 
(e.g., Harrison et al.~2003; Rutten 2020), but this is
beyond the scope of this study.

The content of this paper includes data analysis (Section 2),
a discussion (Section 3), and conclusions (Section 4). 

\section{	DATA ANALYSIS 		}  

When we observe solar emission at {\sl near ultra-violet (NUV)} and
{\sl far ultra-violet (FUV)} wavelengths, we may gather photons from
spicules in plages in the transition region (at formation temperatures
of $T_e \approx 10^4-10^5$). In order to study both 
coherent and incoherent processes, we have to deal with
multiple size distribution functions, including incoherent 
random (Gaussian) noise, as well as coherent avalanche 
processes with power law-like distribution functions, also known as 
``fat-tail'' distribution functions, which occur natually in 
{\sl self-organized criticality (SOC)} systems.

\subsection{	Definitions of Flux Distributions		}

In the following we attempt to model event statistics
with a combination of (i) a Gaussian distribution (originating from
incoherent random processes), and (ii) a power law
distribution, e.g., created by spicular activity in the transition 
region, (Fig.~1). The Gaussian noise is defined in the standard way,
\begin{equation}
	N(F)\ dF = N_0 \exp{\left(-{(F-F_0)^2 \over 2 \sigma_F^2}\right)} \ dF \ ,
\end{equation}
where $F$ is the flux averaged over the duration of an event 
(measured here at a wavelength of 1400 \ang ), $N(F)$ is the histogram 
of observed structures, $F_0$ is the
mean value, $\sigma_F$ is one standard deviation, and
$N_0$ is the normalized number of events.

The second distribution we employ in our analysis is a
power law distribution function, which is defined in the
simplest way by,
\begin{equation}
	N(F)\ dF = N_0 \left({F \over F_0}\right)^{-\alpha_F} \ dF \ ,
\end{equation}
where $\alpha_F$ is the power law slope of the relevant part
of the distribution function.

The flux $F_{\rm IRIS}$ of an {\bf IRIS pixel} is defined by, 
\begin{equation}
	F_{\rm IRIS} = {4\ \pi\ f \ E_{\lambda}\ k \over A\ \Omega} , \quad  
	[{\rm erg}\ {\rm cm}^{-2} {\rm s}^{-1}] 
\end{equation}
where $f$ is the observed flux in [DN] (data number per second), 
$E_{\lambda}$ is the energy of the photon, 
$k$ is the factor that converts the DN to the number of photons, 
$\Omega$ is the pixel size in units of steradians, 
$A$ [cm$^{2}$] is the effective area of IRIS, 
and the unrelated background is subtracted
(Vilangot Nhalil et al.~2020).

\subsection{ 	Pixelized Method of Size Distribution 	}

In this study we use a ``pixelized'' size
distribution method that is more efficient and easier
to calculate than standard size distributions.  
The standard method to sample size distributions $N(F)$ of
SOC avalanches is generally carried out by an algorithm
that detects fluxes of an avalanche event above some given 
threshold $F > F_{thr}$, traces its spatial $A(t)$ and temporal 
evolution $F(t)$, and infers the size of an avalanche
from the spatio-temporal evolution after saturation. 
Such avalanche detections have been accomplished for 
12 IRIS datasets in the study of Vilangot Nhalil 
et al.~(2020). Because the development of an automated 
feature recognition code is a complex and a time-consuming 
task, which needs extensive testing, we explore here a new 
method that is much simpler to apply and requires much 
less data to determine the underlying power law slope 
$\alpha_F$.

We can parameterize a pixelized IRIS image with a Cartesian 
grid, i.e., ${\FF}_{i,j} = {\FF}(x_i,y_j), i=0,...,n_x, j=0,...,n_y$, 
where $n_x$ and $n_y$ are the dimensions of the image,
and $\Delta x = \Delta y$ is the pixel size. 
We can model a 2-D image with a superposition of $n_k$
spatial structures with avalanche areas $A_k$ and 
average fluxes ${\FF}_k$, where the size distributions 
follow a power law distribution, i.e., 
$N({\FF}) \propto {\FF}^{-\alpha_{\FF}}$ (Eq.~2).
The total flux ${\FF}_{\rm tot}$ of such a 2-D distribution, 
which serves here as an analytical model of a 2-D (IRIS) 
image, can then be written as, 
\begin{equation}
	{\FF}_{\rm tot} = \Sigma_{k=1}^{n_k} \  
	{\FF}_k^{-\alpha_{\FF}} \ A_k  \ ,
\end{equation}
where the avalanche areas $A_k$ are required to be
non-overlapping, but area-filling. Areas without
significant avalanche structures, $({\FF} < {\FF}_{thr})$,             
can be included, in order to fulfill flux conservation,
or can be neglected if the flux maximum is much
larger than the threshold value, i.e., ${\FF}_{max} \gg {\FF}_{thr}$. 

In our new method we decompose the flux ${\FF}_k$ and
area $A_k$ of all avalanche components down to the
pixel size level, $\Delta x$. The two requirements
of non-overlapping and area-filling topology yield
a unique mapping of the avalanche number $k$ to the
pixel ranges $i = [i_1(k), i_2(k)]$ and $j = [j_1(k), j_2(k)]$, i.e.,
$k \mapsto i_1(k), ...,i, ...i_2(k)$ and $j_1(k), ...j, ...,j_2(k)$.
For instance, in the case of a rectangular area $A_k$,
the avalanche area $A_k$ is then defined by,
\begin{equation}
	A_k = [i_2(k)-i_1(k)] * [j_2(k)-j_1(k)] \ \Delta x^2
\end{equation}
Adding the areas $A$ and fluxes ${\FF}$ of all $k$ avalanche
components, we obtain then the following total flux 
${\FF}_{\rm tot,pix}$, 
\begin{equation}
	{\FF}_{\rm tot,pix} = \Sigma_{i=1}^{n_x} \Sigma_{j=1}^{n_y} \ 
	{\FF}_{i,j}^{-\alpha_{\FF}} \ \Delta x^2 \ ,
\end{equation} 
which can be set equal to the value of ${\FF}_{\rm tot}$ of the
standard method (Eq.~4) and proves this way that the power
law slopes $\alpha_{\FF}$ of the two methods are identical.
Thus, our new method is parameterized just by a 
different decomposition of elementary components
than in the standard size distribution sampling. 

As a caveat, we have to be aware that the method 
determines size distribution from a single image. 
If the used 2-D image is not representative,
additional 2-D images need to be included. 

The new pixelation 
method is used in the calculations of the values 
$\alpha_{\rm {\FF}_1}$ listed in Table 1 and Fig.~4.

\subsection{	Analysis of IRIS Data 	}

The 12 analyzed 1400 \ang\ SJI images $F(x,y)$ of IRIS are shown in 
Fig.~2, which are identical in time and FOV (field-of-view) with
those of Vilangot Nhalil et al.~(2020), and are also identical
with those used in the study on fractal dimension measurements
(Aschwanden and  Vilangot Nhalil~2022). Note that events
\#6 and \#7; are almost identical, except for a time difference
of 20 min, which can be used for stability tests. 

The 12 IRIS maps shown in Fig.~2 have the following
color code: The Gaussian distribution with values 
$F(x,y) < F_{thr}$ below a threshold of $F_{thr}$ 
is rendered with orange-to-red colors, while the power law
function with the fat-tail $F(x,y) > F_{thr}$ is masked out 
with white color. In other words, all the orange-to-red regions
in the IRIS maps visualize the locations of incoherent random noise
while the white regions
mark the location of SOC-driven coherent avalanches
(probably produced by spicular dynamics in the transition region).
An even crispier representation of the spicular component 
$F(x,y) > F_{thr}$, is displayed with a black-and-white rendering
(Fig.~3), where black depicts locations with power law 
distributions, and white demarcates locations with Gaussian 
distributions.

The information content of an IRIS image can be described
with a 2-D array of flux values $F(x,y)$ at a given time $t$,
or alternatively with a 1-D histogram $N(F)$. 
Since we want to fit a two-component distribution
function (i.e., with a Gaussian and a power law), we need to
introduce a separator between the two distributions,
which we derive from the full width at half maximum (see $F_2$
in Fig.~1). We fit then both distribution functions 
(Eqs.~1 and 2) separately, the Gaussian function in the range
of $[F_1, F_2]$, and the power law function in the range
of $[F_2, F_3]$, as depicted in Fig.~1. The minimum flux ($F_1$)
and maximum flux ($F_3$) are determined from the minimum and
maximum flux value in the image.
We are fitting the distribution functions with a standard 
Gaussian fit method, and with a standard linear regression fit 
for the logarithmic flux function. Note that the power law 
function $N(S)$ appears to be a straight line in a logarithmic 
display only (Fig.~1 bottom panel), i.e., log(N)-log(S), but 
not in a linear representation (Fig.~1 top panel), i.e., 
lin(N)-lin(S), as used here.

The results of the fitting of the observed histograms are 
shown for all 12 datasets in Fig.~(4), where the Gaussian fit 
is rendered with a blue color, and the power law fit with a red color. 
We see that our two-component model for the distribution
function produces accurate fits to the analyzed
IRIS data (histograms in Fig.~4) for 7 datasets
(\# 4-9, 11), while it fails in 5 cases
(\# 1-3, 10, 12). On the other hand, 4 cases
contain sunspots (\# 1-3, 10) and coincide with the
cases with power law fit failures. 

If we would assume that all fluxes are generated by incoherent
random noise, we would not be able to fit the histogrammed 
data at all. Obviously, we would under-predict most of the 
fluxes substantially (blue dashed curves in Fig.~4), which
underscores that the ``fat-tail'' power law function, 
a hallmark of SOC processes, is highly relevant for
fitting the observed IRIS 1400 \ang\ data here. 

In a next step we investigate the numerical values of
the power law slopes $\alpha_F$ of the flux distribution
parameters $F$, which are listed in the third column
of Table 1. At a first glance, it appears that these
values vary wildly in a range of $\alpha_F=0.94$ to
2.13. However, Vilangot Nhalil et al.~(2020) classified
the 12 analyzed datasets into 4 cases containing 
sunspots, and 8 cases containing plages in the transition
region without sunspots. From this bimodal behavior
it was concluded that the power law index of the energy
distribution is larger in plages ($\alpha_E > 2$),
compared with sunspot-dominated active regions 
($\alpha_{E} < 2$), (Vilangot Nhalil et al.~2020). 
In our investigation here, the 4 cases with sunspots
exhibit substantially flatter power law slopes 
$\alpha_F$ (except \#3),
which indicates that sunspot-dominant
distributions are indeed significantly different
from those without sunspots (Table 1). Actually,
we find an even better predictor of this bimodal
behavior, by using the maximum flux $F_{max}$
(Column 6 in Table 1). We find that
flux distributions $N(F) \propto F^{-\alpha_F}$ 
with maximum fluxes less than $F_{max} \lapprox
50$ [DN] exhibit a power law value of
\begin{equation}
	\alpha_F^{obs} \approx 1.70 \pm 0.15,
	\quad  F_{max} < 50\ {\rm DN} \ ,
\end{equation}
which includes the five datasets \#6-9, 11.
In contrast, the seven other datasets \#1-5,
10, 12 have consistently higher maximum
values, $F_{max} \gapprox 50$ DN. 
Instead of using the maximum values $F_{max}$,
we can also use the average fluxes
and find the same bimodal behavior.

Even more significant is that this power law value
is consistent with the theoretical prediction of
the power law slopes (Aschwanden 2012; 2022a;
Aschwanden et al.~2016),
\begin{equation}
	\alpha_{\rm F,SOC} = {9 \over 5} = 1.80 \ .
\end{equation}
Thus we conclude that flux distributions have
a power law slope that agree with the theoretial
prediction under special conditions, such as for small maximum 
fluxes. Moreover we find that magnetic flux distributions 
with sunspots and large magnetic flux imbalances 
produce flatter slopes and failed power law fits 
(Tables 1, 2), see Section 2.4.

\subsection{	HMI Magnetogram Analysis	}

In order to test the universality of the results we repeat
the same analysis for 12 coincident HMI magnetograms onboard 
the {Solar Dynamics Observatory (SDO)}, which
have simultaneous times and identical spatial field-of-views.
The 12 analyzed HMI images are shown in Fig.~5, where
black features indicate negative magnetic polarity, and 
white features indicate positive magnetic polarity. We see sunspots
in at least 4 magnetograms (\#1-3, 10), with two 
sunspots having a negative magnetic polarity (\#1, 2),
and two cases with positive magnetic polarity (\#3, 10).
All 12 magnetograms show mixed polarities, but some are
heavily unbalanced (\#1-5, 10-12). 

We quantify the magnetic flux balance with the ratio $q_{pos}$,
\begin{equation}
	q_{pos} = \left( {\sum_{pos} F_{ij} \over
		\sum_{pos} F_{ij} + |\sum_{neg} F_{ij} |} \right) \ .
\end{equation}
If the magnetic flux (line-of-sight) component is
well-balanced, we would expect a value of $q_{pos}=0.5$,
assuming $\sum_{pos}=|\sum_{neg}|$. 
Only 4 cases have approximately balanced fluxes (\#6-9), 
namely $q_{pos}=[0.44, 0.43, 0.38, 0.44]$, while the other
6 cases have large flux imbalances, from 
$q_{pos}=0.04$ to 0.99 (Table 2 and Fig.~6).
The associated power law slopes of the 4 well-balanced
cases are $\alpha_F=[1.67,1.64,1.79,1.78]=1.72 \pm 0.07$, 
which closely coincide with the theoretical SOC-prediction of
$\alpha_F \approx 1.80$ (Aschwanden 2012; 2022a; 
Aschwanden et al.~2016). 

We analyze the HMI data in the same way as the IRIS data,
by fitting Gaussian distributions (blue curves in Fig.~6)
and power law distribution functions (red curves in Fig.~6),
which clearly show a ``fat-tail'' feature that is far in
excess of the Gaussian function (blue dashed curves in
Fig.~6). We compare the power law slopes $\alpha_F$ obtained
with the two completely different datasets from IRIS and
HMI in Fig.~7, using the ``pixelation'' method. The two
datasets are found to be highly correlated (with
CCC=0.90, if we ignore the outlier \#3). 
Nevertheless, the power law slopes $\alpha_F$ shown in Fig.~7 
are concentrated in two regimes, one that is consistent
with our theoretical SOC prediction of
$\alpha_{\rm F,IRIS} = \alpha_{\rm F,HMI} \approx 1.80$,
while a second cluster is centered around  
$\alpha_{\rm F,IRIS} \approx 1.0-1.5$ and
$\alpha_{\rm F,HMI} \approx 1.0-1.5$ (Fig.~7).
In essence, we find 4 datasets (\# 6-9)
that are consistent with the SOC prediction for
events with well-balanced flux $q_{pos} \approx 0.5$, 
while a second group cannot 
reproduce the SOC model, but can be characterized with
large unbalanced magnetic fluxes (\# 1-5, 10-12).
The flux imbalance, however, is not always decisive.
Tests with variations of the the FOV reveal that the 
arbitrary choice of the FOV (in HMI data) can be more
important in deciding whether the calculated power law 
slope is universally consistent with SOC models
e.g., see event \#11.

The physical interpretation of the HMI data is,
of course, different for the IRIS data. In the
previous analysis of IRIS data we interpreted 
the coherent statistics (in terms of SOC-controlled 
power law functions) due to spicular activity 
in the transition region. In contrast, using the HMI
data, which provides the magnetic field 
line-of-sight component $B_z$,
we can interpret the statistics of incoherent random
distributions in terms of ``salt-and-pepper'' small-scale
magnetic fields in the photosphere, and the coherent 
avalanche statistics in terms of SOC-controlled magnetic 
reconnection processes in nanoflares and larger flares
(Table 3).
Note that the two parameters $\alpha_{\rm F,IRIS}$ amd
$\alpha_{\rm F,HMI}$ are observed independently from
different spacecraft, as well as in markedly
different wavelength bands, i.e., 
$\lambda \approx 1400$ \ang\ for IRIS, and 
$\lambda = 6173$ \ang\ for HMI/SDO magnetograms,
which measures the mean flux $F$ from the
line-of-sight magnetic field component $B_z(x,y)$.
Despite of the very different instruments and
wavelengths, the power law slope $\alpha_F$ 
of the mean flux appears to be universally
valid and consistent with the theoretical 
SOC prediction for datasets with approximate
magnetic flux balance (Fig.~7). However we learned that
the magnetic flux balance and the absence of sunspots
represent additional requirements to warrant the
universality of the SOC slopes. This yields a testable
prediction: If the field-of-view of each HMI magnetogram
is readjusted so that the enclosed magnetic flux becomes
more balanced and no sunspot appears in the FOV, the 
power law slope is expected to approach the theoretical 
universal value of $\alpha_{\rm F,IRIS} \approx 
\alpha_{\rm F,HMI} \approx 1.80$.

\section{	DISCUSSION		} 

In the following we discuss an incoherent random process
(e.g., salt-and-pepper small-scale magnetic elements), 
and two coherent random processes (e.g., spicular dynamics,
and magnetic reconnection), 
which relate to each other as shown in the diagram of Table 3.  

\subsection{	Magnetic Flux Distribution		}

The most extensive statistical study on the size distribution
of magnetic field features on the solar surface has been undertaken
by Parnell et al.~(2009). Combining magnetic field data from
three instruments (SOT/Hinode, MDI/NFI, and MDI/FD on SOHO,
a combined occurrence frequency size distribution was
synthesized that extends over five decades, in the range of 
$\Phi =2 \times 10^{17}-10^{23}$ Mx (Parnell et al.~2009),
\begin{equation}
	N(\Phi ) \propto \left( \Phi_0 \right)^{-1.85\pm 0.14} \quad 
	[{\rm Mx}^{-1} {\rm cm}^{-2}] \ ,
\end{equation}
where the magnetic flux $\Phi$ is obtained from integration
of the magnetic field $B(x,y)$ over a thresholded area 
$A = \int\ dx\ dy$,
\begin{equation}
	\Phi = \int B(x,y)\ dx\ dy \quad [{\rm Mx}]\ .
\end{equation}
If we equate the magnetic flux $\Phi$ with the
mean flux $F$ of an event in standard SOC models, we predict
a power law slope of (Aschwanden 2012; 2022a; Aschwanden et al.~2016),
using $d=3$, $D_V=5/2$, and $\gamma=1$,
\begin{equation}
        \alpha_{\rm F,SOC} = 1 + {(d-1) \over (\gamma D_V)} = {9 \over 5}
        = 1.80 \ ,
\end{equation}
which agree well with the result 
(Eq.~10) observed by Parnell et al.~(2009).
A lower value was found from cellular automaton simulations,
$N(\Phi)\approx \Phi^{-1.5\pm0.05}$ (Fragos et al.~2004),
where flux emergence is driven by a percolation rule,
similar to the percolation model of Seiden and Wentzel (1996),
or Balke et al.~(1993).
Mathematical models have been developed to model the percolation
phenomenon, based on combinatorial and statistical concepts
of connectedness that exhibit universality in form of
powerlaw distributions.

\subsection{	Universality of SOC Size Distributions  	}

Power law-like size distributions are the hallmark
of self-organized criticality systems. Statistical
studies in the past have collected SOC parameters
such as length scales $L$, time scales $T$, peak flux
rates $P$, mean fluxes $F$, fluences and energies
$E=F\times T$, mono-fractal and multi-fractal dimensions
(Mandelbrot 1977), in order to test whether the theoretically
expected power law size distributions, or the power law
slopes of waiting times, agree with the observed
distributions (mostly observed in astrophysical systems). 
The universality of SOC models 
(Aschwanden 2012; 2022a; Aschwanden et al.~2016) 
is based on four scaling laws:
the scale-free probability conjecture
$N(L) \propto L^{-d}$, classical diffusion
$L\propto T^{\beta/2}$, the flux-volume relationship
$F \propto V^{\gamma}$, and the Euclidean 
scaling law, $P \propto L^{\gamma d}$, where
$d=3$ is the Euclidean dimension, $\beta \approx 1$ is the
classical diffusion coefficient, $\gamma \approx 1$ the
flux-volume proportionality, while $D_A=3/2$ and $D_V=5/2$
are the mean fractal dimensions in 2-D and 3-D Euclidean
space. The standard SOC model is expressed in terms of
these universal constants: $d=3$, $\gamma=1$, $\beta=1$. 
Consequently, the four basic scaling laws reduce to 
$N(L) \propto L^{-3}$, $L \propto T^{1/2}$,  
$F \propto L^{2.5}$, and $P \propto L^3$. Since we measure the
mean flux $F$ in this study, our main test of the
universality of SOC models if formulated in terms of
the flux-volume relationship $F \propto V^{\gamma}$, 
leading to the 
power law slope $\alpha_{\rm F,SOC}=1.80$ (Eq.~12).

The SOC-inferred scaling laws hold for a large number 
of phenomena. This implies that our SOC formalism is 
universal in the sense that the statistical size distributions 
are identical for each phenomenon, displaying a universal
power law slope of $\alpha_{\rm F,SOC}=1.80$.   
When we conclude that the power law slope $\alpha_F$ is universal,
the SOC model implies that the flux-volume proportionality
($\gamma \approx 1$) as well as the mean fractal dimension 
($d=3$, $D_V \approx 2.5$) are universal too. 

\subsection{	Phenomena with SOC 	}

Once we establish the self-consistency of power law slopes 
between theoretical (SOC) and observed size distributions, the
next question is what physical processes are at work. 
We envision four different types of phenomena (Table 3): 
(i) Gaussian random noise in IRIS data);
(ii) spicular plage events in the transition region 
(described by the power law size distribution in IRIS data);
(iii) salt-and-pepper small-scale magnetic structures
(described by the random noise distributions in HMI magnetograms);
and (iv) magnetic reconnection processes in flares and nanoflares
(described by the power law size distribution in HMI data).
However, there are deviations from these rules. We found that
the power law 
distributions are modified in the presence of sunspots, when 
the magnetic flux is unbalanced, or when the FOV is
arbitrarily chosen. Under ideal conditions, 
the SOC scaling laws are fulfilled universally, independent
of the wavelength or plasma temperature.
Magnetic field data (from HMI/SDO) or $\lambda \approx 1400$ \ang\ 
(from IRIS) appear to produce emission in volumes that are proportional 
in the photosphere or transition zone, even when they are formed at 
quite different temperatures, i.e., $T_{\rm phot} \approx 5800$ K 
in the photosphere and $T_{\rm TR} \approx 10^4-10^5$ K in the 
transition region.  

Another ingredient of the SOC model is the scale-free probability
conjecture, i.e., $N(L) \propto L^{-d} = L^{-3}$, which
cannot be uniquely linked to a particular  physical process. Parnell
et al.~(2009) conclude that a combination of emergence, 
coalescence, cancellation, and fragmentation may possibly produce
power law size distributions of spatial scales $L$. 
Alternative models include the
turbulence and the Weibull distributions (Parnell 2002). 
Munoz-Jaramillo et al.~(2015) study the best-fitting distribution
functions for 11 different databases of sunspot areas, sunspot
group areas, sunspot umbral areas, and magnetic fluxes. 
They find that a linear combination of Weibull and log-normal
distributions fit the data best (Munoz-Jaramillo et al.~2015). 
Weibull and log-normal distributions combine two distribution
functions, similar to our synthesis of a Gaussian-plus-power-law 
distribution.

A general physical scenario of a
power law size distribution is the evolution of
avalanches by exponential growth (Rosner and Vaiana 1978), 
with subsequent saturation (logistic growth) after a random 
time interval, which produces an exact power law function  
(Aschwanden et al.~1998). Our approach to model the size
distribution of solar phenomena with two different functions,
employing a Gaussian noise and a power law tail, reflects the
duality of incoherent and coherent random 
components, in both the data from IRIS and HMI (Table 3). In summary, 
incoherent random components include salt-and-pepper 
small-scale magnetic features, while coherent components
include spicular avalanches, and magnetic reconnection
avalanches from nanoflares to large flares. 

\subsection{	Granular Dynamics	}

The physical understanding of solar (or stellar) granulation
has been advanced by numerical magneto-convection models and
N-body dynamic simulations, which predict the evolution of
small-scale (granules) into large-scale
features (meso- or super-granulation), organized by surface
flows that sweep up small-scale structures and form clusters of
recurrent and stable granular features (Hathaway et al.~2000;
Berrilli et al.~1998, 2005; Rieutord et al.~2008, 2010;
Cheung and Isobe 2014; Martinez-Sykora et al.~2008).
An analytical model of convection-driven generation of
ubiquitous coronal waves is considered in Aschwanden et al.~(2018b).
The fractal multi-scale dynamics has been found to be operational
in the Quiet-Sun photosphere, in quiescent non-flaring states, as
well as during flares (Uritsky et al.~2007, 2013; 
Uritsky and Davila 2012).
The fractal structure of the solar granulation is obviously a
self-organizing pattern that is created by a combination of
subphotospheric magneto-convection and surface flows, which are
turbulence-type phenomena. 

The interpretation of granulation as the cause of the 
Gaussian "noise" in IRIS data is controversial for two
reasons: (i) {\bf The intensity measured by IRIS 1400 in 
non-magnetic areas has densities that originate from the
middle chromosphere, rather than from the underlying photosphere.}
(ii) No convective signal propagates to these heights and 
densities, and thus the scale of granulation cannot be
probed at these heights (Martinez-Sykora et al 2015).

\subsection{	Spicular Dynamics			}

One prominent feature in the transition region is the phenomenon
of {\sl ``moss''}, which appears as a bright dynamic pattern with
dark inclusions, on spatial scales of $L \approx 1-3$ Mm, which
has been interpreted as the upper transition region above active
region plages, and below relatively hot loops 
(De Pontieu et al.~1999; 2014).
Our measurement of structures in the IRIS 1400 \ang\ channel
is sensitive to a temperature range of $T_e \approx 10^4-10^5$ K,
and thus is likely to include chromospheric and transition region
phenomena such as: spicules II (De Pontieu et al.~2007), 
macro-spicules, dark mottles, dynamic fibrils, surges, miniature 
filament eruptions, etc. Theoretical models include 
the rebound shock model (Sterling and Hollweg 1988), 
pressure-pulses in the high atmosphere (Singh et al.~2019), 
Alfv\'enic resonances (Sterling 1998),
magnetic reconnection models for type II spicules (De Pontieu et al.~2007),
ion-neutral collisional damping (De Pontieu 1999), 
leakage of global p-mode oscillations (De Pontieu et al.~2004), 
MHD kink waves (Zaqarashvili and Erdelyi 2009), 
vortical flow models (Kitiashvili et al.~2013), and
magneto-convective driving by shock waves (De Pontieu et al.~2007). 

The fact that we obtain a power law size distribution 
($\alpha_F=1.70\pm0.15$, Table 1), which is very similar 
to solar flares in general, $\alpha_{\rm F,SOC}=1.80$,
implies the universality of the SOC framework. Furthermore
we find power law-like size distributions for spicular events,
rather than a Gaussian distribution, which tells us that 
spicule events need to be modeled in terms of SOC-driven 
avalanches, instead of Gaussian random distributions.
As mentioned above, the difference between incoherent 
and coherent random processes is the following: Gaussian 
statistics reflect the operation of a memoryless stationary 
random process; while avalanche processes such as occurring 
in SOC systems are characterized by extended spatial and 
temporal correlations, i.e., the unfolding of an avalanche 
is influenced by the imprint of earlier avalanches on 
the system.

We propose that spicules around magnetic elements
are responsible for the power law slope $\alpha_F$ observed
in those areas. This appears to be a plausible interpretation
since these dynamical phenomena are very relevant in
plage and network areas. For instance, event \#11 shows
a fully unbalanced magnetic configuration, which supports
the idea that strong magnetic fields, fragmented in
small-scale elements in plage and/or network seems to be
the relevant characteristics, rather than flux balance
over an arbitrary FOV.

\subsection{	Salt-and-Pepper Magnetic Field	        }

We interepret the random noise Gaussian distribution of
magnetic fluxes in Quiet-Sun regions as small-scale
magnetic field ``pepper-and-salt'' structures, also
called {\sl magnetic carpet} (Priest et al.~2002), 
where the black and white
color in magnetograms (Fig.~5) corresponds to negative and
positive polarity. 
The fact that we obtain two distinctly different 
size distributions (Gaussian vs. power law) indicates
at least two different physical mechanisms, one being
an incoherent random (Gaussian) process, the other one
being a coherent (power law-like) avalanche process.
The salt-and-pepper structure is generated apparently
by an incoherent random process, rather than by a
coherent avalanching process, according to our fits. 
This may constrain the origin of the solar magnetic field, 
being created by emergence, submergence, coalescence, 
cancellation, fragmentation, and/or small-scale dynamos, 
etc. Not all would be expected to yield Gaussian 
statistics (e.g., fragmentation processses often yield 
log-normal distributions; Verbeeck et al.~2019). 

\subsection{	Magnetic Reconnection 		}	

The re-arrangement of the stress-induced solar 
magnetic field requires ubiquitous and permanent 
(but intermittent) magnetic reconnection processes 
on all spatial and temporal scales.
Our study finds power law size distributions,
with a slope of $\alpha_F=1.72\pm 0.07$ from HMI 
magnetograms, which is similar to flares in general
(see Aschwanden et al.~2016 for a review of all 
wavelengths (e.g., gamma-rays, hard X-rays, soft X-rays, 
UV, EUV, FUV, etc). This tells us that there is a strong 
correlation between the photospheric field (in HMI images) 
and the transition region (in IRIS images), as evident 
from the cross-correlation coefficient of CCC=0.90 
shown in Fig.~7. The fractal multi-scale 
dynamics apparently operates in the quiet photosphere, 
in the quiescent non-flaring state, as well as during 
flares in active regions (Uritsky and Davila 2012).

\section{	CONCLUSIONS			}

Solar and stellar flares, pulsar glitches, auroras,
lunar craters, as well as earthquakes, landslides,
wildfires, snow avalanches, and sandpile avalanches
are all driven by self-organized criticality (SOC),
which predicts power law-like occurrence 
frequency (size) distributions and waiting time 
distribution functions. What is new in our studies
of SOC systems is that we are now able to calculate 
the slope $\alpha_x$ of power law functions, which 
allows us to test SOC models by comparing the
observed (and fitted) distribution functions with
the theoretically predicted values. In this study
we compare statistical distributions of SOC parameters
from different wavelengths and different instruments
(UV emission observed with IRIS and magnetograms
with HMI). The results of our study are:

\begin{enumerate}

\item{The histogrammed distribution of fluxes $N(F)$
obtained from an IRIS 1400 \ang\ image, or from a HMI 
magnetogram, cannot be fitted solely by a Gaussian 
function, but requires a two-component function, 
such as a combination of a Gaussian and a power law
function, a ``fat-tail'' extension above 
some threshold. We define a separator between
the two functions above the full width at half maximum. 
We obtain power law slopes of $\alpha_F=1.70\pm 0.15$
from the IRIS data, and $\alpha_F=1.72\pm 0.07$
from the HMI data, which agree with the theoretical
SOC prediction of $\alpha_F=1.80$, and thus 
demonstrate universality across UV wavelengths and
magnetograms. Moreover, it agrees with the five
order of magnitude extending power law distribution
sampled by Parnell et al.~(2009), $\alpha_F 
= 1.85\pm0.14$.}

\item{Tables 1 and 2 show the following characterizations 
of the 12 selected datasets: 
4 cases with sunspots,  
5 cases that have a max flux $<50$ DN, 
4 cases with a magnetic flux balance of $q_{pos} \approx 0.4$, and 
5 cases that agree with the theortical prediction $a_F=1.8$ 
(see values flagged with YES/NO in Tables 1 and 2).
In summary, the universality of the flux power law slope
($a_F=1.80$) depends on the absence of sunspots,
small maximum fluxes, magnetic flux balance,
and the choice of the field-of-view of an active region.
In other words, the scale-free probability inherent to
SOC models requires some special conditions for
magnetic field parameters. 
Strong magnetic fields, fragmented in small-scale elements 
in plage and/or network seems to be
the relevant characteristics, rather than flux balance
over an arbitrary FOV.}

\item{We designed an algorithm that produces
``pixelized'' size distributions from a single image 
(e.g., from a UV image or a magnetogram). In this
method, the flux and area of each avalanche
event is decomposed down to the pixel size level,
which allows us to calculate the power law slope of
the size flux distibution without requiring an
automated feature recognition code.
The method is computationally
very fast and does not require any particular
automated pattern recognition code.}

\item{We can characterize the analyzed size 
distributions in terms of four distinctly different 
physical interpretations: (i) the Gaussian
random noise distribution in IRIS data;
(ii) spicular plage events in the transition region 
(described by the power law size distribution in IRIS data);
(iii) salt-and-pepper small-scale magnetic structures
(described by the random noise distributions in HMI magnetograms);
and (iv) magnetic reconnection processes in flares and nanoflares
(described by the power law size distribution in HMI data).}

\end{enumerate}

Future work may include: (i) Testing of the SOC-predicted
size distributions with power law slopes $\alpha_F$ for
all available (mean) fluxes $F$ (in HXR, SXR, EUV, etc.); 
(ii) testing the selection of different FOV sizes in 
the absence or existence of sunspots, and magnetic 
flux balance; (iii) and cross-comparing the
``pixelization'' method with the standard method.
Ultimately these methods should help us
to converge the numerical values in SOC models.

\medskip
\acknowledgments
{\sl Acknowledgements:}
We acknowledge constructive and insightful comments of 
two reviewers and stimulating discussions (in alphabetical order), with 
Paul Charbonneau, Adam Kowalski, Karel Schrijver, and Vadim Uritsky. 
This work was partially supported by NASA contract NNX11A099G
``Self-organized criticality in solar physics'', NASA contract
NNG04EA00C of the SDO/AIA instrument, and the IRIS contract 
NNG09FA40C to LMSAL.
IRIS is a NASA small explorer mission developed and operated by 
LMSAL with mission operations executed at NASA Ames Research 
Center and major contributions to downlink communications 
funded by ESA and the Norwegian Space Centre.

\clearpage

\def\ref#1{\par\noindent\hangindent1cm {#1}}

\section*{	References	}

\ref{Aschwanden, M.J., Dennis, B.R., and Benz, A.O. 1998,
 	{\sl Logistic avalanche processes, elementary time structures, 
	and frequency distributions in solar flares},
 	ApJ 497, 972}
\ref{Aschwanden, M.J. 2011,
        {\sl Self-Organized Criticality in Astrophysics. The Statistics
        of Nonlinear Processes in the Universe}, ISBN 978-3-642-15000-5,
        Springer-Praxis: New York, 416p.}
\ref{Aschwanden, M.J. 2012,
        {\sl A statistical fractal-diffusive avalanche model of a
        slowly-driven self-organized criticality system},
        A\&A 539, A2, (15 p)}
\ref{Aschwanden, M.J. 2014,
        {\sl A macroscopic description of self-organized systems and
        astrophysical applications}, ApJ 782, 54}
\ref{Aschwanden,M.J., Crosby,N., Dimitropoulou,M., Georgoulis,M.K.,
        Hergarten,S., McAteer,J., Milovanov,A., Mineshige,S., Morales,L.,
        Nishizuka,N., Pruessner,G., Sanchez,R., Sharma,S., Strugarek,A.,
        and Uritsky, V. 2016,
        {\sl 25 Years of Self-Organized Criticality: Solar and Astrophysics}
        Space Science Reviews 198, 47-166.}
\ref{Aschwanden, M.J., Gosic, M., Hurlburt, N.E., and Scullion, E. 2018b, 
 	{\sl Convection-driven generation of ubiquitous coronal waves}, 
	ApJ 866, 72 (13pp).}
\ref{Aschwanden, M.J. 2022a,
	{\sl The fractality and size distributions of astrophysical
	self-organized criticality systems},
	ApJ 934:33}
\ref{Aschwanden, M.J. 2022b,
	{\sl Reconciling power-law slopes in solar flare and nanoflare
	size distributions}, ApJL 934:L3}
\ref{Aschwanden, M.J. and Vilangot Nhalil, N. 2022,
	{\sl Interface region imaging spectrograph (IRIS) observations 
	of the fractal dimension in the solar atmosphere},
	Frontiers in Astronomy and Space Sciences, Manuscript ID 999329}
\ref{Balke,A.C., Schrijver, C.J., Zwaan,C., and Tarbell,T.D. 1993,
        {\sl Percolation theory and the geometry of photospheric
        magnetic flux concentrations}, Solar Phys. 143, 215.}
\ref{Bak, P., Tang, C., and Wiesenfeld, K. 1987,
        {\sl Self-organized criticality: An explanation of 1/f noise},
        Physical Review Lett. 59(27), 381}
\ref{Bak, P., Tang, C., and Wiesenfeld, K. 1988,
        {\sl Self-organized criticality},
        Physical Rev. A 38(1), 364}
\ref{Bak, P. 1996,
        {\sl How Nature Works. The Science of Self-Organized Criticality},
        New York: Copernicus}
\ref{Berrilli, F., Florio, A., and Ermolli, I. 1998,
	{\sl On the geometrical properties of the chromospheric network},
	Sol.Phys. 180, 29-45}
\ref{Berrilli, F., Del Moro, D., Russo, S., et al. 2005,
        {\sl Spatial clustering of photospheric structures},
        ApJ 632, 677}
\ref{Burlaga, L.F. and Lazarus, A.J. 2000, 
 	{\sl Lognormal distributions and spectra of solar wind plasma 
	fluctuations: Wind 1995-1998},
 	JGR 105, 2357}
\ref{Ceva, H. and Luzuriaga, J. 1998,
 	{\sl Correlations in the sand pile model: From the log-normal 
	distribution to self-organized criticality},
 	Physics Letters A 250, 275}
\ref{Cheung, M.C.M. and Isobe, H. 2014,
	{\sl Flux emergence (Theory)},
	LRSP 11, 3.}
\ref{De Pontieu, B. 1999,
 	{\sl Numerical simulations of spicules driven by weakly-damped 
	Alfv\'en waves I. WKB approach},
 	A\&A 347, 696}
\ref{De Pontieu, B., Berger, T.E., Schrijver, C.J., and Title, A.M. 1999,
	{\sl Dynamics of transition region 'moss' at high time resolution}.
	Sol.Phys. 190, 419}
\ref{De Pontieu, B., Erdelyi, R., and James, S.P. 2004,
 	{\sl Solar chromospheric spicules from the leakage of photospheric 
	oscillations and flows},
 	Nature 430, 536}
\ref{De Pontieu,B., McIntosh,S., Hansteen,V.H., Carlsson,M., 
	Schrijver,C.J., et al. 2007,
 	{\sl A tale of two spicules: The impact of spicules 
	on the magnetic chromosphere},
 	PASJ 59, S655}
\ref{De Pontieu, B., Title, A.M., Lemen, J.R., Kushner, G.D., 
	Akin, D.J., Allard, B., Berger, T., Boerner, P., 2014,
	{The Interface Region Imaging Spectrograph (IRIS)},
	Sol.Phys. 289, 2733}
\ref{Fontenla, J.M., Curdt, W., Avrett, E.H., and Harder, J. 2007, 
 	{\sl Log-normal intensity distribution of the quiet-Sun FUV 
	continuum observed by SUMER},
 	A\&A 468, 695}
\ref{Fragos, T., Rantsiou, E., and Vlahos, L. 2004,
 	{\sl On the distribution of magnetic energy storage in solar 
	active regions},
 	A\&A 420, 719.}
\ref{Gallagher, P.T., Phillips, K.J.H., Harra-Murnion, L.K., et al.
        1998, {\sl Properties of the Quiet Sun EUV network},
        A\&A 335, 733}
\ref{Giles, D.E., Feng,H., and Godwin,R.T. 2011,
 	{\sl On the bias of the maximum likelihood estimator for the 
	two-parameter Lomax distribution},
 	Econometrics Workshop Paper EWP1104, ISSN 1485}
\ref{Harrison, R.A., Harra, L.K., Brkovic, A., and Parnell, C.E. 2003,
 	{\sl A Study of the unification of quiet-Sun transient-event 
	phenomena},
 	A\&A 409, 755}
\ref{Hathaway, D.H., Beck, J.G., Bogart, R.S., et al. 2000,
        {\sl The photospheric convection spectrum}
        SoPh 193, 299}     
\ref{Hosking, J.R.M. and Wallis, J.R. 1987,
 	{\sl Parameter and quantile estimation for the generalized 
	Pareto distribution},
 	Technometrics, 29(3), 339}
\ref{Hudson, H.S. 1991,
	{\sl Solar flares, microflares, nanoflares, and coronal heating},
	Sol.Phys. 133, 357}
\ref{Jensen, H.J. 1998, 
 	{\sl Self-Organized Criticality: Emergent complex behaviour in 
	physical and biological systems},
 	Cambridge University Press.}
\ref{Katsukawa, Y. and Tsuneta, S. 2001,
 	{\sl Small fluctuation of coronal X-ray intensity and a
	signature of nanoflares},
 	ApJ 557, 343}
\ref{Kitiashvili, I.N., Kosovichev, A.G., Lele, S.K., Mansour, N.N.
	and Wray, A.A. 2013,
	{\sl Ubiquitous solar eruptions driven by magnetized
	vortex tubes},
	ApJ 770, 37}
\ref{Krucker, S. and Benz, A.O. 1998,
	{\sl Energy distribution of heating processes in the quiet
	solar corona},
	ApJ 501, L213}
\ref{Kunjaya, C., Mahasena, P., Vierdayanti, K., and Herlie, S. 2011,
	{\sl Can self-organized critical accretion disks generate a log-normal 
	emission variability in AGN?},
 	ApSS 336, 455}
\ref{Lomax, K.S. 1954,
	J. Am. Stat. Assoc. 49, 847}
\ref{Lu, E.T. and Hamilton, R.J. 1991,
 	{\sl Avalanches and the distribution of solar flares}, 
	ApJ 380, L89}
\ref{Mandelbrot, B.B. 1977, {\sl The Fractal Geometry of Nature}.
        W.H.Freeman and Company: New York}
\ref{Martinez-Sykora, J., Hansteen,V., and Carlsson, M. 2008,
 	{\sl Twisted Flux Tube Emergence From the Convection Zone to 
	the Corona}, ApJ 679, 871}
\ref{Martinez-Sykora,J., Rouppe van der Voort, L., Carlsson, M., 
	De Pontieu, B., Pereira, T.M.D., Boerner, P., Hurlburt, N., et al. 2015,
 	{\sl Internetwork Chromospheric Bright Grains Observed With 
	IRIS and SST}
	ApJ 803, 44}
\ref{McAteer,R.T.J., Aschwanden,M.J., Dimitropoulou,M., Georgoulis,M.K.,
        Pruessner, G., Morales, L., Ireland, J., and Abramenko,V. 2016,
        {\sl 25 Years of Self-Organized Criticality: Numerical Detection Methods},
        SSRv 198 217-266.}
\ref{Mitzenmacher, M. 2004,
 	{\sl A brief history for generative models for power law and 
	lognormal distributions},
 	Internet mathematics 1(2), 226}
\ref{Munoz-Jaramillo, A., Senkpeil, R.R., Windmueller, J.C., Amouzou, E.C., et al. 2015,
 	{\sl Small-scale and Global Dynamos and the Area and Flux Distributions 
	of Active Regions, Sunspot Groups, and Sunspots: A Multi-database Study}, 
 	ApJ 800, 48}
\ref{Parnell, C.E. 2002,
	{\sl Nature of the magnetic carpet - 1. Distribution of
	magnetic fluxes},
	MNRAS 335/2, 398.}
\ref{Parnell, C.E., DeForest, C.E., Hagenaar, H.J., Johnston, B.A., 
	Lamb, D.A., and Welsch, B.T. 2009, 
 	{\sl A Power-Law Distribution of Solar Magnetic Fields Over More 
	Than Five Decades in Flux},
	ApJ 698, 75-82}
\ref{Priest,E.R., Heyvaerts, J.F., and Title, A.M. 2002,
 	{\sl A Flux-Tube Tectonics Model for Solar Coronal Heating 
	Driven by the Magnetic Carpet},
 	ApJ 576, 533}
\ref{Rathore, B. and Carlsson, M. 2015, 
	{\sl The Formation of IRIS Diagnostics. VI. The Diagnostic Potential of the 
	C II Lines at 133.5 nm in the Solar Atmosphere},
	ApJ 811, 80}
\ref{Rathore, B., Carlsson, M., Leenaarts, J., De Pontieu B. 2015, 
	{\sl  The Formation of Iris Diagnostics. VIII. Iris Observations in the 
	C II 133.5 nm Multiplet},
	ApJ 811, 81}
\ref{Rieutord, M., Meunier, N., Roudier, T., et al. 2008,
        {\sl Solar super-granulation revealed by granule tracking},
        A\&A 479, L17}
\ref{Rieutord, M., Roudier, T., Rincon, F., 2010,
        {\sl On the power spectrum of solar surface flows},
        A\&A 512, A4}
\ref{Rosner, R., and Vaiana, G.S. 1978,
 	{\sl Cosmic flare transients: constraints upon models for energy storage 
	and release derived from the event frequency distribution},
 	ApJ 222, 1104}
\ref{Rutten et al. 2020,
	{\sl SoHO campfires in SDO images}.
	arXiv:2009.00376v1, astro-ph.SR.}
\ref{Scargle, J.D. 2020,
 	{\sl Studies in astronomical time-series analysis. VII. An enquiry 
	concerning nonlinearity, the rms-mean flux relation, and lognormal flux 
	distributions},
 	ApJ 895, 90}
\ref{Schrijver, C.J., Hagenaar, H.J., and Title, A.M.  1997
 	{\sl On the patterns of the solar granulation and super-granulation},
 	ApJ 475, 328-337}
\ref{Seiden, P.E. and Wentzel, D.G. 1996,
 	{\sl Solar active regions as a percolation phenomenon II.}
 	ApJ 460, 522}
\ref{Singh, B., Sharma, K., and Srivastava, A.K. 2019,
	{\sl On modelling the kinematics and evolutionary properties
	of pressure pulse-driven impulsive solar jets},
	Ann.Geophys. 37, 891}
\ref{Sterling,A.C. and Hollweg,J.V. 1988,
 	{\sl The rebound shock model for solar spicules: Dynamics at long 
	times}
 	ApJ 327, 950}
\ref{Sterling, A.C. 1998,
	{\sl Alfv\'enic resonances on ultraviolet spicules},
	ApJ 508, 916}
\ref{Uritsky, V.M., Paczuski, M., Davila, J.M., and Jones, S.I. 2007,
        {\sl Coexistence of self-organized criticality and intermittent
        turbulence in the solar corona},
        Phys.~Rev.~Lett. 99(2), id. 025001}
\ref{Uritsky, V.M., and Davila, J.M. 2012,
 	{\sl Multiscale Dynamics of Solar Magnetic Structures},
 	ApJ 748, 60}
\ref{Uritsky, V.M., Davila, J.M., Ofman, L., and Coyner, A.J. 2013,
        {\sl Stochastic coupling of solar photosphere and corona},
        ApJ 769, 62}
\ref{Verbeeck, C., Kraaikamp, E., Ryan, D.F., and Podladchikova, O. 2019,
 	{\sl Solar Flare Distributions: Lognormal Instead of Power Law?},
 	ApJ 884, 50}
\ref{Vilangot Nhalil, N.V., Nelson, C.J., Mathioudakis, M., and Doyle, G.J. 2020,
 	{\sl Power-law energy distributions of small-scale impulsive events 
	on the active Sun: results from IRIS},
 	MNRAS 499, 1385}
\ref{Warren, H.P., Reep, J.W., Crump, N.A., and Simoes, P.J.A. 2016,
	{\sl Transition region and chromospheric signatures of	
	impulsive heating events. I. Observations},
	ApJ 829:35}
\ref{Weibull, W. 1951,
	{\sl A statistical distribution function of wide applicability},
	J. Appl. Mech 18(3) 293}
\ref{Zaqarashvili, T.V. and Erdelyi, R. 2009,
	{\sl Oscillations and waves in solar spicules},
	SSRv 149, 355}
\clearpage


\begin{table}
\begin{center}
\small	
\caption{Results of 12 datasets obtained with IRIS 1400 \ang\ : 
the power law slope $\alpha_F$ of the flux distribution,
the separator flux $F_2$, and the maximum flux $F_{max}$.
Note that the power law slope $\alpha_F$ agrees with the theoretical
prediction of $\alpha_F=9/5= 1.8$ in 5 cases, whenever there is no
sunspot and the maximum flux $F_{max}$ amounts to less than
a critical value of $F_{max} \lapprox 50$ DN. The values 
$\alpha_F$ in parenthesis are ignored in the calculation of the
averages (second-last line).} 
\medskip
\begin{tabular}{ccccccc}
\hline
\hline
Number	& Phenomenon 	& Power law 	   & Agrees with                  & Separator  & Maximum   & Max.flux   \\
Dataset	& 1400 \ang     & slope fit        & prediction                   & flux       & flux      & criterion  \\
IRIS	&		& $\alpha_{\rm F}$ & $\alpha_{\rm F} \approx 1.8$ & $F_2$      & $F_{max}$ & $<50$ DN   \\
\#	&		&  	           &                              & [DN ]      & [DN]      &            \\
\hline
\hline
1       & Sunspot       & (1.51$\pm$0.04)  & NO                           &  21        & 121       & NO   \\
2       & Sunspot       & (1.23$\pm$0.02)  & NO                           &  32        & 190       & NO   \\
3       & Sunspot       & (2.13$\pm$0.06)  & NO                           & 128        & 243       & NO   \\
4       & Plage         & (0.94$\pm$0.02)  & NO                           &  20        & 108       & NO   \\
5       & Plage         & (1.02$\pm$0.01)  & NO                           &  36        & 199       & NO   \\
6       & Plage         &  1.59$\pm$0.02   & YES                          &  17        &  50       & YES  \\
7       & Plage         &  1.59$\pm$0.03   & YES                          &   9        &  26       & YES  \\
8       & Plage         &  1.92$\pm$0.03   & YES                          &   8        &  28       & YES  \\
9       & Plage         &  1.81$\pm$0.01   & YES                          &  13        &  42       & YES  \\
10      & Sunspot       & (1.25$\pm$0.02)  & NO                           & 120        & 501       & NO   \\
11      & Plage         &  1.61$\pm$0.02   & YES                          &   9        &  31       & YES  \\
12      & Plage         & (1.40$\pm$0.05)  & NO                           &  22        &  54       & NO   \\
\hline
        & Observations  & 1.70$\pm$0.15    &                              &            &           &      \\       
        & Theory        & 1.80             &                              &            &           &      \\
\hline
\end{tabular}
\end{center}
\end{table}


\begin{table}
\begin{center}
\small	
\caption{Results of 12 datasets obtained with HMI/SDO, showing
the power law slope $\alpha_F$ of the flux distribution,
the separator flux $F_2$, the magnetic flux balance $q_{pos}$,  
the magnetic field strength $B_z$, the magnetic flux balance $q_{pos}$,
and the fractal dimension $D_A$.
Note that the power law slope $\alpha_F$ agrees with the theoretical
prediction of $\alpha_F=9/5\approx 1.8$ in 5 cases approximately,
when there is no sunspot and the magnetic flux is balanced. The values of
$\alpha_F$ in parenthesis are ignored in the calculation of
the averages.} 
\medskip
\begin{tabular}{ccccccccc}
\hline
Number	& Phenomenon 	& Power law 	& Matching	  & Separator            & Magnetic  & Magnetic     & Matching & Fractal \\
Dataset	&  		& slope fit     & prediction      & flux                 & field     & flux balance & balance  & dimension \\
HMI	&		& $\alpha_{\rm F}$ & $\alpha_{\rm F}\approx 1.8$ & $F_2$ & $B_z$     & $q_{pos} $   & $q_{pos}\approx 0.50$ & $D_A$\\

\#	&		&  	        &                 & [DN]      & [G]      &        &           \\
\hline
\hline
1       & Sunspot       & (1.32$\pm$0.03) & NO            & 8           & +1073 & (0.04) & NO  & 1.54 \\
2       & Sunspot       & (1.27$\pm$0.01) & NO            & 6           & -1729 & (0.16) & NO  & 1.55 \\
3       & Sunspot       & (0.92$\pm$0.02) & NO            & 5           & -2076 & (0.99) & NO  & 1.59 \\
4       & Plage         & (1.32$\pm$0.01) & NO            & 5           & +1785 & (0.29) & NO  & 1.58 \\
5       & Plage         & (1.33$\pm$0.02) & NO            & 5           & -1186 & (0.81) & NO  & 1.57 \\
6       & Plage         &  1.67$\pm$0.02  & YES           & 4           & +1854 & 0.44 & YES & 1.51 \\
7       & Plage         &  1.64$\pm$0.02  & YES           & 4           & -1011 & 0.43 & YES & 1.51 \\
8       & Plage         &  1.79$\pm$0.03  & YES           & 4           & -1022 & 0.38 & YES & 1.49 \\
9       & Plage         &  1.78$\pm$0.03  & YES           & 5           &  +955 & 0.44 & YES & 1.50 \\
10      & Sunspot       & (0.94$\pm$0.01) & NO            & 7           & -1055 & (0.34) & NO  & 1.66 \\
11      & Plage         &  1.72$\pm$0.02  & YES           & 5           & +2058 & (0.92) & NO  & 1.51 \\
12      & Plage         & (1.22$\pm$0.03) & NO            & 4           & +1036 & (0.88) & NO  & 1.52 \\
\hline
        & Observations  & 1.72$\pm$0.07   &               &             & & 0.42 $\pm$ 0.03 &   & 1.54$\pm$0.05 \\       
        & Theory        & 1.80            &               &             & & 0.50            &   & 1.50  \\
\hline
\end{tabular}
\end{center}
\end{table}


\begin{table}
\begin{center}
\normalsize
\caption{Diagram of phenomena observed with different instruments
(IRIS, HMI), different wavelengths (columns), for incoherent and coherent 
random processes (rows).}
\medskip
\begin{tabular}{|l|c|c|}
\hline
\hline
 		     	 & IRIS		& HMI	        	  \\
 		     	 & 1400 \ang\ \ & 6173 \ang\ 		  \\
\hline
incoherent random process & ?  	 & salt-and-pepper 	  \\
(Gaussian function)  	 &               & small-scale magnetic fields \\
\hline
{coherent} random process & spicules & flares, nanoflares      \\
(power law function) 	 &		 & magnetic reconnection   \\
\hline
\end{tabular}
\end{center}
\end{table}

\begin{figure}
\centerline{\includegraphics[width=1.0\textwidth]{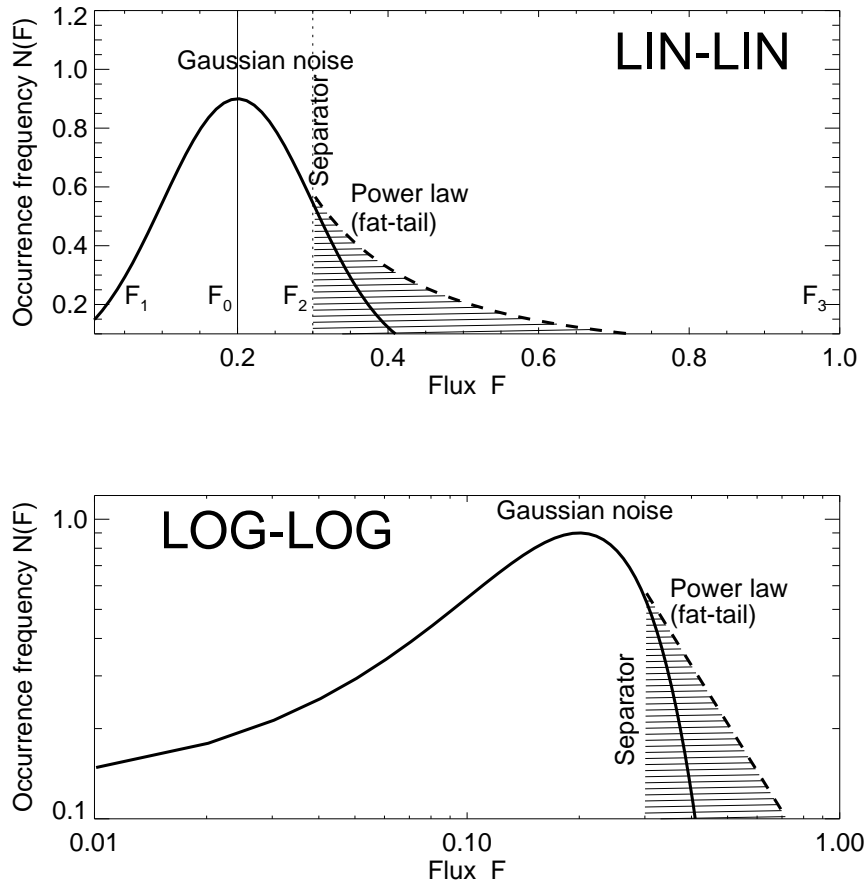}}
\caption{A schematic of the two size distributions is shown:
a Gaussian function for the incoherent random statistics, 
and a power law function (also called fat-tail) for the 
statistics of coherent avalanche events, separated at 
a critical value $F_2$. The upper panel shows a linear
(LIN-LIN) representation, the lower panel a logarithmic
(LOG-LOG) representation.} 
\end{figure}

\begin{figure}
\centerline{\includegraphics[width=1.0\textwidth]{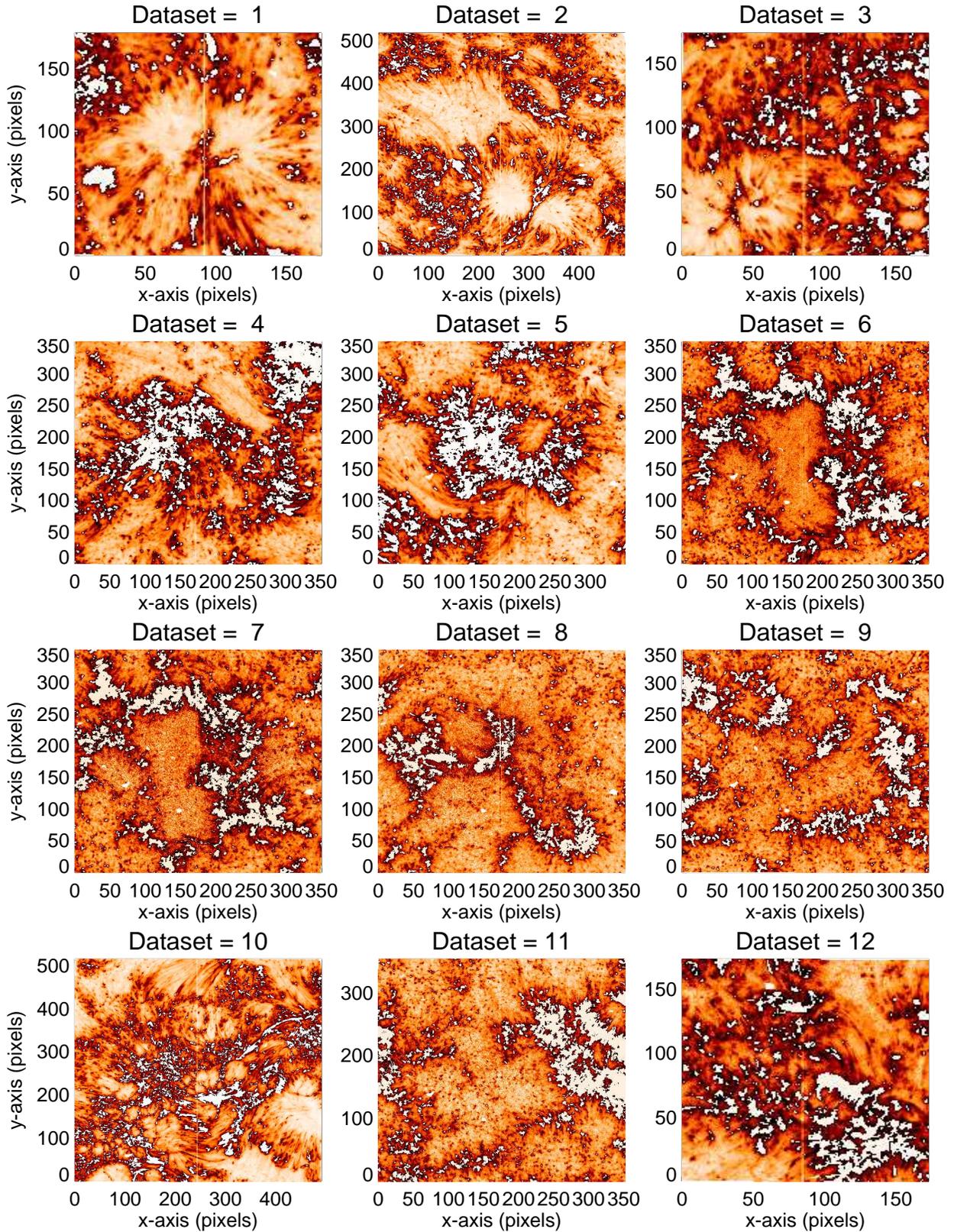}}
\caption{Intensity maps of 12 different active regions
and Quiet-Sun regions, observed with IRIS SJI 1400 \ang . 
Gaussian random noise is rendered in orange-to-red color, while 
spicules and network cells are masked out with white color.}
\end{figure}

\begin{figure}
\centerline{\includegraphics[width=0.9\textwidth]{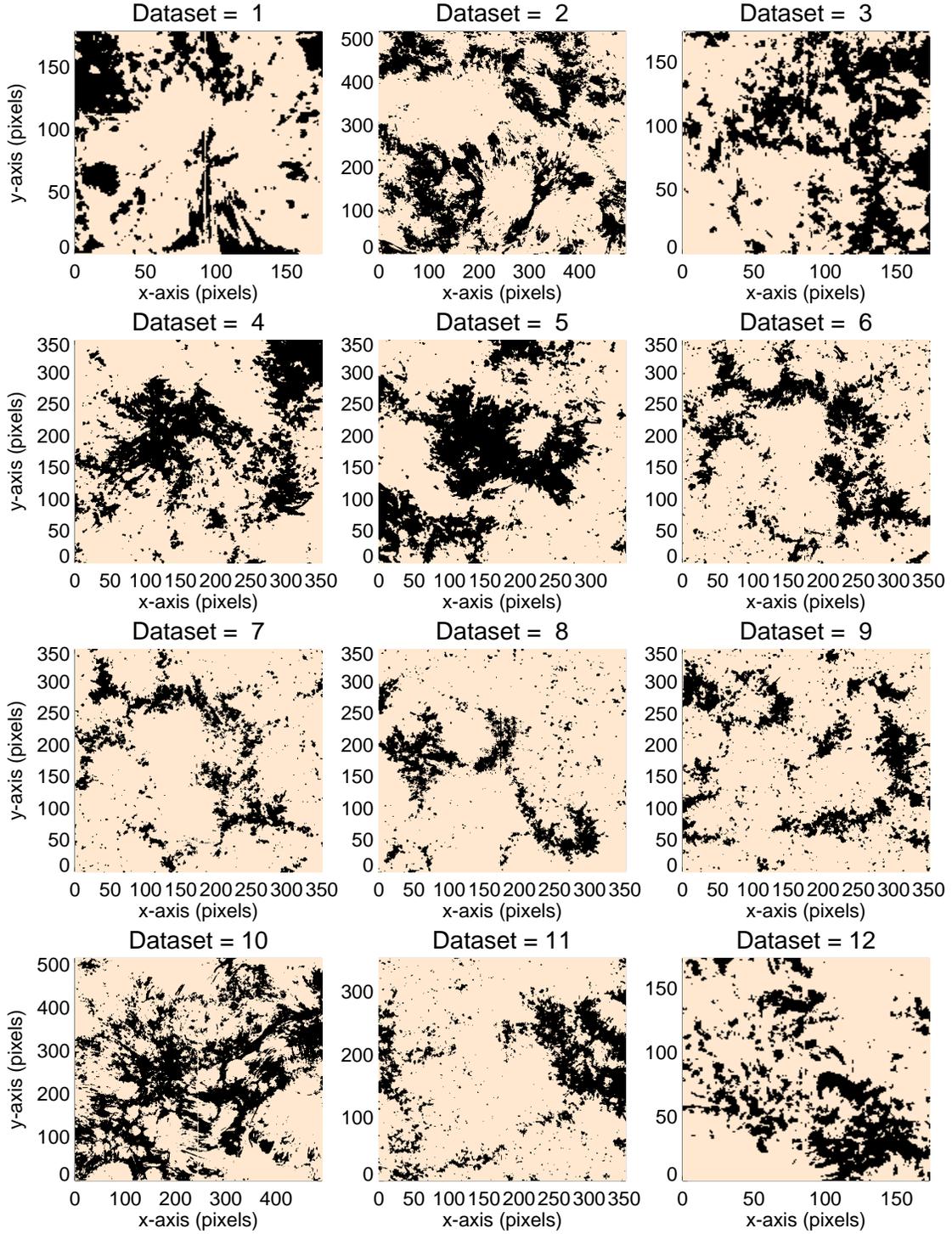}}
\caption{Intensity maps of 12 different active regions
and Quiet-Sun regions, observed with IRIS SJI 1400 \ang . 
Gaussian random noise is masked out (with peak fluxes 
$F(x,y) < F_{thr}$), while network cells and spicules 
are rendered in black.}
\end{figure}

\begin{figure}
\centerline{\includegraphics[width=0.9\textwidth]{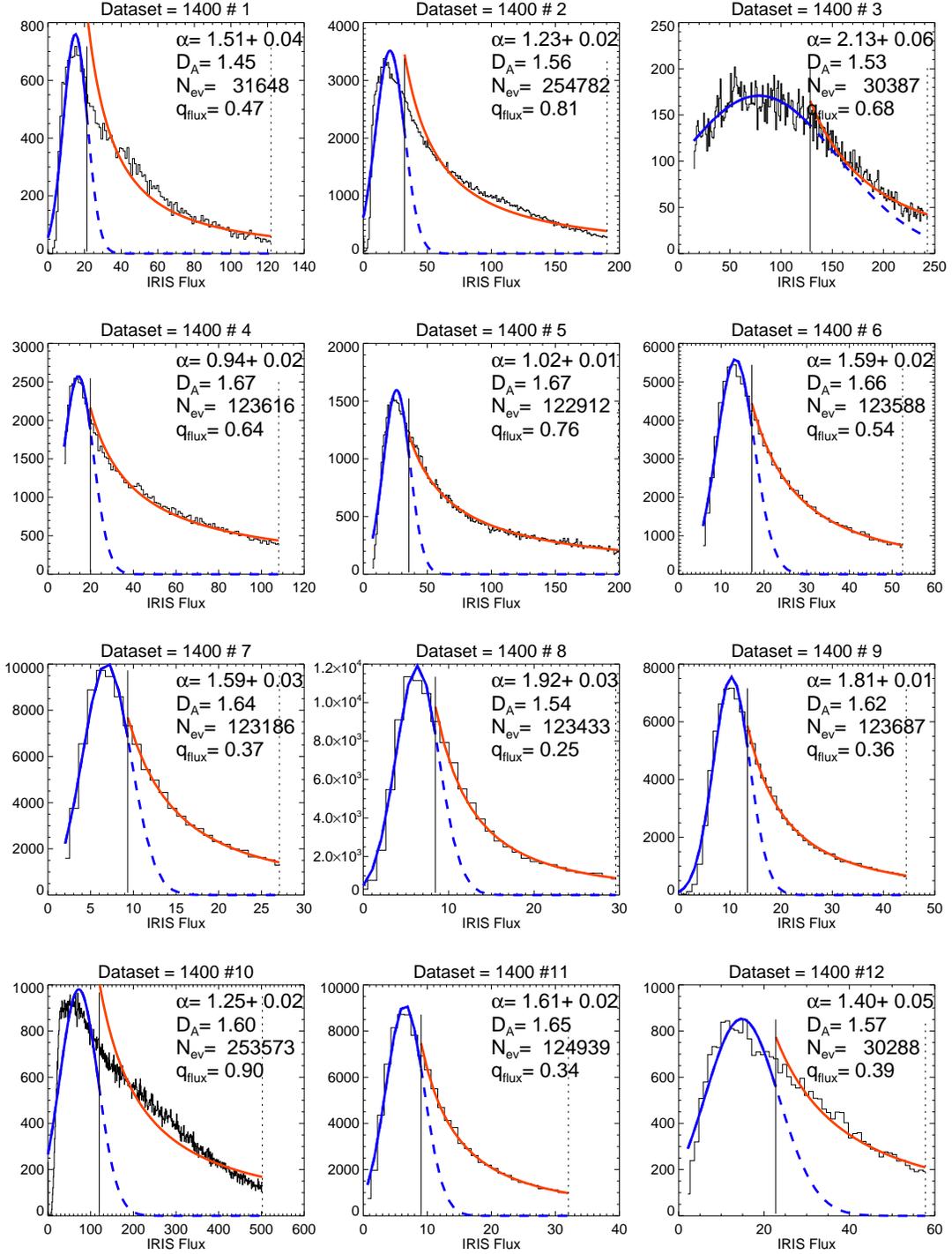}}
\caption{Flux histograms of 12 different regions in plages
of transition regions, observed with IRIS SJI 1400 \ang\ . 
The flux distribution of granules is fitted with a Gaussian
function (blue curve, $F < F_2$), and extrapolated
with dashed blue curves. The flux distribution of spicules is 
fitted with a power law distribution function (thick red 
curve. The separation of the two distributions at $F_2$ 
is marked with a vertical thin line.}
\end{figure}

\begin{figure}
\centerline{\includegraphics[width=0.9\textwidth]{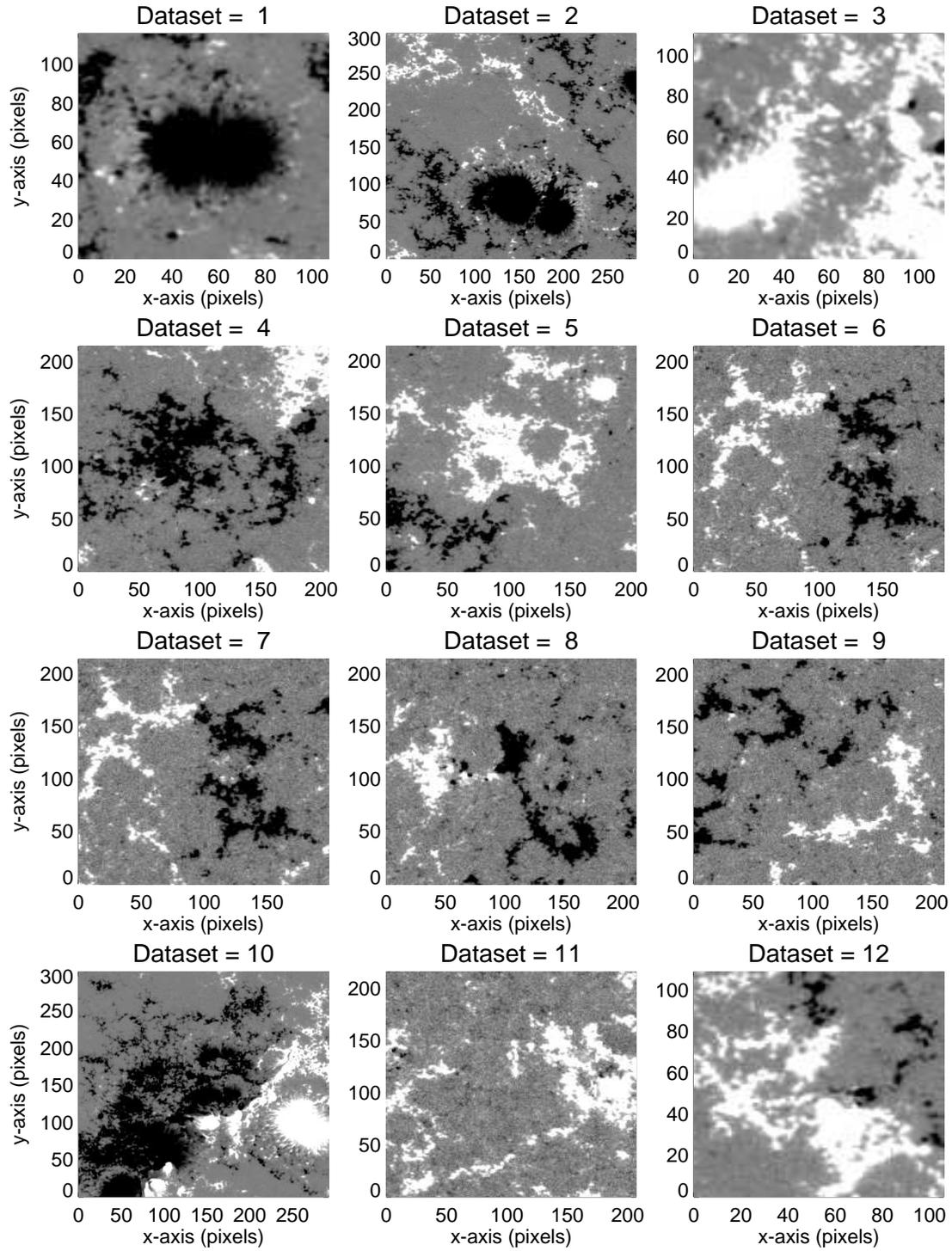}}
\caption{Magnetograms of 12 different active regions
and plage regions, observed with HMI/SDO . The black 
color indicates negative magnetic polarity, and the 
white color represents positive magnetic polarity.} 
\end{figure}

\begin{figure}
\centerline{\includegraphics[width=0.9\textwidth]{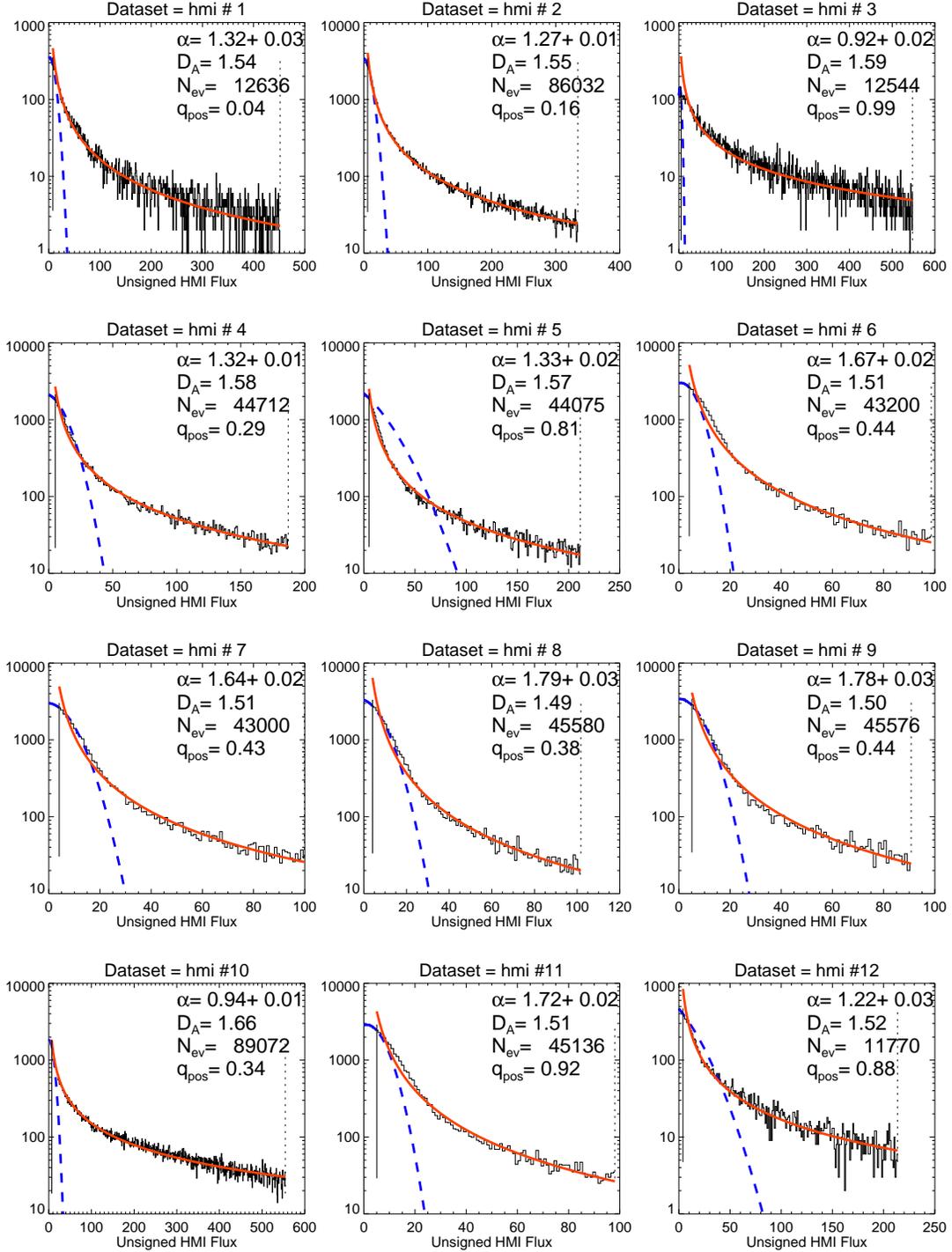}}
\caption{Histograms of different solar regions, 
observed in magnetograms with HMI/SDO. The size
distribution of salt-and-pepper magnetic noise is fitted
with a Gaussian function (blue curve), the extrapolation
of the Gaussian (dashed blue curve), while the 
distribution of magnetic features are fitted with 
power law functions (red curves).}
\end{figure}

\begin{figure}
\centerline{\includegraphics[width=0.9\textwidth]{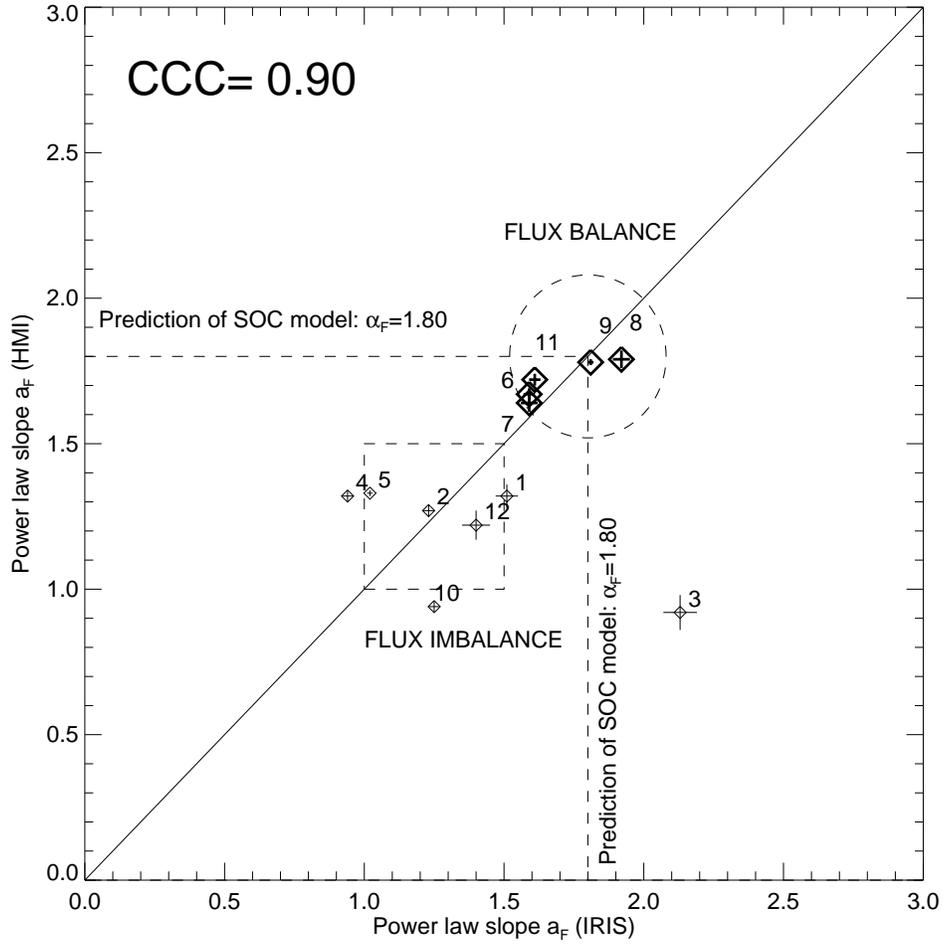}}
\caption{The power law slopes $\alpha_F$ are calculated 
for 12 datasets for two independent instruments and wavelengths: 
from IRIS data (x-axis) and from HMI/SDO data (y-axis). 
Note that 5 datasets (\#6-9, 11) coincide 
approximately with the theoretically expected value 
$\alpha_F=1.80$ (marked with a circle). The other 7 cases
(shown in rectangle) 
are subject to sunspots, relatively large peak fluxes,
and large magnetic flux imbalances.}
\end{figure}

\end{document}